\documentstyle[prl,aps,epsf,epsfig,array]{revtex}

\newcommand{\bi}{\begin{itemize}}
\newcommand{\ei}{\end{itemize}}
\newcommand{\bc}{\begin{center}}
\newcommand{\ec}{\end{center}}

\newcommand{\be}{\begin{equation}}
\newcommand{\ee}{\end{equation}}

\newcommand{\bqn}{\begin{eqnarray}}
\newcommand{\eqn}{\end{eqnarray}}

\begin{document}

\title{Reply to "Comment on One loop renormalization of soliton quantum mass 
corrections in (1+1)-dimensional scalar field theory models"
(hep-th/0211149)}

\author{G. Flores-Hidalgo \thanks{E-mail:gflores@cbpf.br}}

\address{{\it  Centro Brasileiro de Pesquisas Fisicas-CBPF,
Rua Dr. Xavier Sigaud 150, 22290-180 Rio de Janeiro, RJ, Brazil}}

\maketitle

\begin{abstract}
I agree with the authors of hep-th/0211149 that the claim made in
Phys.Lett. B{\bf542}, 282 (2002) is incorrect and that the derivation of its
main formula, although correct, contains two compensating errors. In this 
reply the main formula of Phys.Lett. B{\bf542}, 282 (2002) is rederived.
This new derivation shows that not only the energy momentum cut off regularization
method still works in the calculation of the soliton quantum mass 
corrections, but also that the so called mode number regularization
emerges naturally from it. 
\vspace{0.34cm}
\noindent
\end{abstract}
\vspace{1cm}

In Ref. \cite{rebhan1} the authors refuted the claim 
made in Ref. \cite{flores1}, that previous works on the one
loop soliton quantum mass corrections did not include
a surface term from a partial integration. I agree with the authors
of Ref. \cite{rebhan1}. As the authors of this reference pointed out,
in the derivation of the main
formula of Ref. \cite{flores1} I used incorrectly
the asymptotic behaviour of the continuous phase shift and also
there was a wrong counting of the density of states. These two
mistakes compensate in such a way that the correct formula, Eq. (19)
of Ref. \cite{flores1}, was obtained. The way I derived
that equation, "showed" that the final result is independent of the
regularization method. On the other hand, the authors of 
Ref. \cite{rebhan2} showed that the energy momentum cut off regularization 
method gave incorrect results when it was used in the calculation 
of the soliton quantum mas corrections. Therefore, in Ref. \cite{flores1}
I concluded that the claim above was wrong. 
In this reply I rederive the formula given by Eq. (19) of Ref. \cite{flores1},
using the energy momentum cut off regularization method, showing that it gives 
the correct result for the soliton quantum mass corrections.
Then my position about the claim made in Ref. \cite{rebhan2} remains
to be the same.

In the notation of Ref. \cite{flores1}, the bare one
loop soliton quantum mass correction is given by

\be
\Delta M_{bare}=\frac{1}{2}\sum_{i}\omega_i+\frac{1}{2}\sum_{q}\omega(q)
-\frac{1}{2}\sum_{k}\omega^0(k)\;,
\label{e1}
\ee
where $i$ and $q$ label the eigenfrequencies of respectively
the discrete and continuous modes in the presence of the soliton,
whereas $k$ labels those of the vacuum modes. Since $\omega(q)=
\omega(-q)=\sqrt{q^2+m^2}$ and $\omega^0(k)=\omega^0(-k)=\sqrt{k^2+m^2}$
I can write Eq. (\ref{e1}) as

\be
\Delta M_{bare}=\frac{1}{2}\sum_{i}\omega_i+\sum_{q=0}
^{\infty}\omega(q)-\sum_{k=0}^{\infty}\omega^0(k)\;.
\label{a1}
\ee
The continuous sums in the above equation are quadratically divergent
and generally the difference of these is logarithmically divergent. Then
these expressions need to be regularized. As an alternative to the energy
momentum cut off regularization method, the authors of Ref. \cite{rebhan2}
introduced the so called mode number regularization method. Below
I will show that not only the energy momentum cut off regularization method
gives correct results in the calculation of the soliton quantum mass
corrections but also, that the so called mode number regularization
method emerges naturally from the former. The idea
is to use the same cut off $\Lambda$ in the sums over $q$ and $k$
in Eq. (\ref{a1}), then

\be
\Delta M_{bare}=\frac{1}{2}\sum_{i}\omega_i+\sum_{q=0}
^{\Lambda}\omega(q)-\sum_{k=0}^{\Lambda}\omega^0(k)\;.
\label{a2}
\ee

To write the continuous sums in Eq. (\ref{a2}) in an integral form, I
enclose the system in a box of finite size $L$, impose periodic boundary 
conditions and finally let $L\to\infty$. For the free modes I obtain
$k_n=2\pi n/L$, $n=0,\pm 1,\pm 2,...$, from which the density 
of states for the free modes $L/(2\pi)$ is obtained. On the other hand 
I have

\be
q_n=\frac{2\pi n}{L}-\frac{\delta(q_n)}{L},~~n=0,\pm1,\pm2,..\;.
\label{ad3}
\ee
The phase shift $\delta(q)$, can be chosen in two ways. The customary in 
scattering
theory is to choose it (for physical reasons) in such a way that
$\delta(\pm\infty)=0$. In this case the phase shift is discontinuous
(with discontinuity at the origin) and I denote it as $\delta_D(q)$. 
The other way is to choose it in such
a way that $\delta(0)=0$ and in this case it is a continuous function and
I denote it as $\delta_C(q)$. I expect that the result for
Eq. (\ref{a2}) be independent of the phase shift being continuous or 
discontinuous. For large momentum, $\delta_D(q)$ behaves like

\be
\delta_D(\Lambda)=-\frac{\langle V\rangle}{2\Lambda}\;,
\label{a4}
\ee
where $\langle V\rangle$ is given by Eq. (15) of Ref. \cite{flores1}. Also
I have that $\delta_D(0^+)={\cal N}\pi$, where ${\cal N}$
is the number of discrete eigenfrequencies. On the other hand
for $\delta_C(q)$ I have

\be
\delta_C(\Lambda)=-{\cal N}\pi-\frac{\langle V\rangle}{2\Lambda}\;.
\label{a5}
\ee
The derivatives of the continuous and discrete phase shifts are the same,
and I denote them in both cases as $d\delta(k)/dk$. 

In terms of discretized eigenfrequencies, Eq. (\ref{a2}) can be
written as

\be
\Delta M_{bare}=\frac{1}{2}\sum_{i=1}^{\cal N}\omega_i+\sum_{n=n_0}
^{N'}\sqrt{q_n^2+m^2}-\sum_{n=0}^{N}\sqrt{k_n^2+m^2}\;,
\label{a6}
\ee
where in the second term the sum starts at $n=n_0$ not necessarily equal
to zero and ends at $n=N'$ not necessarily equal to $N$, 
since I require that in both sums, $q_n$ and $k_n$ ranges, according 
to Eq. (\ref{a2}),
from zero to the same cut off $\Lambda$. 
For the free eigenfrequencies $N$ is given in terms of the cut off 
$\Lambda$ by, 

\be
\Lambda=2\pi N/L\;.
\label{aa1}
\ee 
To find $n_0$ and $N'$ I consider separately the continuous and discontinuous 
phase shifts.\\\

{{\bf\it Discontinuous phase shift}}~~In this case, setting $q_{n_0}=0^+$ in
Eq. (\ref{ad3}) and using $\delta_D(0^+)={\cal N}\pi$, I have
$2\pi n_0/L-{\cal N \pi}/L=0$, from which I obtain $n_0={\cal N}/2$.
On the other hand setting $\Lambda=q_{N'}$ in Eq. (\ref{ad3}) and 
using $\delta_D(\Lambda)\to 0$ I obtain $\Lambda=2\pi N'/L$
and comparing with Eq. (\ref{aa1}) I get $N'=N$. I have to call 
attention to the fact that the solution $n_0={\cal N}/2$, only makes sense
when ${\cal N}$ is even. For odd ${\cal N}$ it is obtained a half integer value for $n_0$,
a solution not present in the spectrum as given by Eq. (\ref{ad3}). Of course
such a solution does not satisfies the periodic boundary conditions \footnote{It is
easy to see that it satisfies anti periodic boundary conditions}. Then, in this
last case I have to drop out the term $n_0={\cal N}/2$ from the sum in the second
term of Eq. (\ref{a6}), the sum beginning at the next integer larger than 
${\cal N}/2$. If I write 
${\cal N}=2{\cal M}+1$, this means that the sum must begin in 
$n={\cal M}+1$. Also in the last sum of Eq. (\ref{a6}) the contribution from
the zero momentum, $n=0$, must be multiplied by a factor $1/2$, since in the
original sum given by Eq. (\ref{e1}) the corresponding term is counted 
only once. To proceed with my derivation I consider first, the case 
${\cal N}=2{\cal M}$ even. In this case $n_0={\cal M}$ and Eq. (\ref{a6}) can
be written as 
  
\bqn
\Delta M_{bare}&=&\frac{1}{2}\sum_{i=1}^{{\cal N}}\omega_i+
\sum_{n={\cal M}}^N\sqrt{q_n^2+m^2}-\sum_{n=0}^N\sqrt{k_n^2+m^2}
\nonumber\\
&=&\frac{1}{2}\sum_{i=1}^{{\cal N}}\omega_i
-\sum_{n=0}^{{\cal M}-1}\sqrt{k_n^2+m^2}+
\sum_{n=0}^N\left(\sqrt{q_n^2+m^2}-\sqrt{k_n^2+m^2}\right)\;,
\label{s1}
\eqn

From Eq. (\ref{ad3}) I obtain

\be
\sqrt{q_n^2+m^2}=\sqrt{k_n^2+m^2}-\frac{\delta_D(k_n)}{L
\sqrt{k_n^2+m^2}}+{\cal O}\left(L^{-2}\right)\;.
\label{s2}
\ee
Using Eq. (\ref{s2}), $\delta_D(0^+)=2{\cal M}\pi$ and Eq. (\ref{a4}) in Eq. 
(\ref{s1}), and taking $L\to\infty$ I get

\bqn
\Delta M_{bare}&=&\frac{1}{2}\sum_{i=1}^{{\cal N}}\omega_i
-{\cal M}m-\frac{\omega(k)}{2\pi}\delta_D(k)\left.\right|_{0^+}^{\Lambda}
+\int_{0}^{\Lambda}\frac{dk}{2\pi}\omega(k)
\frac{d}{dk}\delta(k)\nonumber\\
&=&\frac{1}{2}\sum_{i=1}^{\cal N}\omega_i
+\frac{\langle V \rangle}{4\pi}+
\frac{1}{2}\int_{-\Lambda}^{\Lambda}\frac{dk}{2\pi}\omega(k)
\frac{d}{dk}\delta(k)+{\cal O}\left(\Lambda^{-1}\right)\;.
\label{ff2}
\eqn
Then, subtracting the tadpole graph contribution, as given by Eq. (14)
of Ref. \cite{flores1},

\be
\frac{\langle V\rangle}{4}\int_{-\Lambda}^{\Lambda}\frac{dk}{2\pi}
\frac{1}{\sqrt{k^2+m^2}}
\label{tgra}
\ee
from Eq. (\ref{ff2}) and taking $\Lambda\to\infty$ I obtain the main formula, 
Eq. (19) of Ref. \cite{flores1}.

In the case in which ${\cal N}=2{\cal M}+1$, as explained above, I have to multiply
the $n=0$ contribution for the free eigenfrequencies with a factor $1/2$
and the sum over the eigenfrequencies in the presence of the soliton must begin in 
$n={\cal M}+1$.  Then, Eq. (\ref{a6}) can be written as

\bqn
\Delta M_{bare}&=&\frac{1}{2}\sum_{i=1}^{{\cal N}}\omega_i+
\sum_{n={\cal M}+1}^N\sqrt{k_n^2+m^2}-\sum_{n=1}^N\sqrt{q_n^2+m^2}
-\frac{m}{2}\nonumber\\
&=&\frac{1}{2}\sum_{i=1}^{{\cal N}}\omega_i-\sum_{n=1}^{{\cal M}}\sqrt{k_n^2+m^2}
-\frac{m}{2}
+\sum_{n=1}\left(\sqrt{k_n^2+m^2}-\sqrt{q_n^2+m^2}\right)\;.
\label{xx1}
\eqn
Using Eq. (\ref{s2}), $\delta_D(0^+)=(2{\cal M}+1)\pi$ and Eq. (\ref{a4}) in 
Eq. (\ref{xx1}) and taking $L\to\infty$ I obtain

\bqn
\Delta M_{bare}&=&\frac{1}{2}\sum_{i=1}^{{\cal N}}\omega_i
-\left({\cal M}+\frac{1}{2}\right)m-\left.\frac{\omega(k)}{2\pi}\delta_D(k)
 \right|_{0}^{\Lambda}+\int_{0}^{\Lambda}\frac{dk}{2\pi}\omega(k)
 \frac{d}{dk}\delta(k)\nonumber\\
 &=&\frac{1}{2}\sum_{i=1}^{{\cal N}}\omega_i+\frac{\langle V\rangle}{4\pi}
 +\frac{1}{2}\int_{-\Lambda}^{\Lambda}\frac{dk}{2\pi}\omega(k)\frac{d}{dk}
 \delta(k)+{\cal O}\left(\Lambda^{-1}\right)\;.
\label{xx2}
\eqn
Again, subtracting the tadpole graph contribution, given by expression 
(\ref{tgra}), from Eq. (\ref{xx2}) I recover the main formula of Ref. 
\cite{flores1}.\\\

{\it continuous phase shift}~~ In this case, since $\delta_C(0)=0$, from Eq. 
(\ref{ad3}) I find $n_0=0$. On the other hand for the upper limit I have,
from Eqs. (\ref{ad3}) and (\ref{a5})
\be
\Lambda=\frac{2N'\pi}{L}+\frac{{\cal N}\pi}{L}\;.
\label{a7}
\ee
From Eqs. (\ref{aa1}) and (\ref{a7}) I find that $N'=N-{\cal N}/2$. In order
to check that also in this case the main formula of Ref. \cite{flores1} is obtained
I consider for simplicity only the case in which ${\cal N}=2{\cal M}$. 
In this case Eq. (\ref{a6}) can be written as 
\bqn
\Delta M_{bare}&=&\frac{1}{2}\sum_{i=1}^{{\cal N}}\omega_i+
\sum_{n=0}^{N-{\cal M}}\sqrt{q_n^2+m^2}-\sum_{n=0}^N\sqrt{k_n^2+m^2}\nonumber\\
&=&\frac{1}{2}\sum_{i=1}^{{\cal N}}\omega_i-
\sum_{n=0}^{{\cal M}-1}\sqrt{q_{n-{\cal M}}^2
+m^2}+\sum_{n=0}^{N}\left(\sqrt{q_{n-{\cal M}}^2
+m^2}-\sqrt{k_n^2+m^2}\right)\;.
\label{e7}
\eqn
From Eq. (\ref{ad3}) I obtain

\be
\sqrt{q_{n-{\cal M}}^2+m^2}=\sqrt{k_n^2+m^2}-\frac{k_n\left(\delta_C(k_n)+
2{\cal M}\pi
\right)}{L\sqrt{k_n^2+m^2}}+{\cal O}\left(L^{-2}\right)\;.
\label{aded1}
\ee
Using Eq. (\ref{aded1}), $\delta_C(0)=0$ and Eq. (\ref{a5}) in Eq. (\ref{e7}),
and taking $L\to\infty$ I obtain

\bqn
\Delta M_{bare}&=&\frac{1}{2}\sum_{i=1}^{\cal N}\omega_i-m{\cal M}
-\frac{\omega(k)}{2\pi}\left(\delta_C(k)+2{\cal M}\pi\right)\left.\right|_{0}^
{\Lambda}+\frac{1}{2}\int_{-\Lambda}^{\Lambda}\frac{dk}{2\pi}
\omega(k)\frac{d}{dk}\delta(k)\nonumber\\
&=&\frac{1}{2}\sum_{i=1}^{{\cal N}}\omega_i+
\frac{\langle V \rangle}{4\pi}+
\frac{1}{2}\int_{-\Lambda}^{\Lambda}\frac{dk}{2\pi}
\omega(k)\frac{d}{dk}\delta(k)+{\cal O}\left(\Lambda^{-1}\right)\;.
\label{ff1}
\eqn
Subtracting from the above expression the tadpole graph contribution as
given by expression (\ref{tgra}) and taking $\Lambda\to\infty$,
I obtain again Eq. (19) of Ref. \cite{flores1}. In the case in which ${\cal N}$
is odd, it is easy to show that also in this case it is recovered the main
formula of Ref. \cite{flores1}. 

It should be noted from Eqs. (\ref{s1}), (\ref{e7}) or (\ref{xx1}) that
the discretized sum over the eigenfrequencies in the presence of the
soliton contains ${\cal N}$ terms
less than the sum over the free eigenfrequencies, being ${\cal N}$, the number
of discrete eigenfrequencies in the presence of the soliton. In the so called 
mode number regularization \cite{rebhan2}
this fact is used as the starting point. Here I have shown that it follows
naturally from using adequately an energy momentum cut off regulator
for the divergent sums in Eq. (\ref{e1}). In Ref. \cite{rebhan2}, in the
case of the sine-Gordon model, in which there is only one discrete eigenvalue,
it is not very clear why it is excluded the term with $n=0$ from the sum over the
eigenfrequencies in the presence of the soliton. Here I have shown that such a term 
needs to be excluded since it does not satisfies the imposed periodic boundary 
conditions.

In Ref. \cite{rebhan1} the authors also claim that the solitons in the field 
theoretical model, with density potential given by $\phi^2\cos^2{\rm Ln}(\phi^2)$, 
only exist
semiclassically. The model $\phi^2\cos^2{\rm Ln}(\phi^2)$ was
introduced in Ref. \cite{flores2}. The argument used by the authors of Ref. 
\cite{rebhan1} is a one loop calculation performed in Ref.
\cite{rajaraman}. In this reference the authors showed that the solitons
of the triple degenerate $\phi^6$ model only exist at the classical level, 
at the quantum level an infrared divergence arises that can not be renormalized. 
This occurs because the potential of the Schrodinger equation for the
eigenfrequencies in the presence of the soliton, takes different values 
at $x\to -\infty$ and $x\to\infty$, that is, because the second derivative
of the density potential at the neighbor minima are different. In the model,
$\phi^2\cos^2{\rm Ln}(\phi^2)$ the second derivative in all minima are equal,
but the third derivative in the neighbor minima are no longer equal. Since
two loops quantum corrections are proportional to the third derivative of
the density potential, the authors of Ref. \cite{rebhan1}
concluded trivially that at two loop order, something similar to what 
happens in the one loop calculation of Ref. \cite{rajaraman}
also will occur in the model $\phi^2\cos^2{\rm Ln}(\phi^2)$. In a future
work I will show that this is not the case.

\vspace{1cm}

{\large \bf Acknowledgments}\\
I would like to thank N. F. Svaiter for useful discussions and A. P. C. Malbouisson
for reading the manuscript. Also a grant from 
Conselho Nacional de Desenvolvimento Cient\'\i fico e Tecnol\'ogico 
(CNPq-Brazil) is acknowledged.

\end{document}